\documentclass[runinaddress,showpacs,twocolumn,prl]{revtex4}
\usepackage{amssymb}
\usepackage{amsmath}
\usepackage{graphicx}

\input{tcilatex}

\begin{document}

\title{Finite size effects of helical edge states in HgTe/CdTe quantum wells}
\date{\today}
\author{Bin Zhou$^{1,2}$, Hai-Zhou Lu$^{1}$,\ Rui-Lin Chu$^{1},$ Shun-Qing
Shen$^{1}$, and Qian Niu$^{3}$}
\affiliation{$^{1}$Department of Physics, and Center for Theoretical and Computational
Physics, The University of Hong Kong, Pokfulam Road, Hong Kong, China\\
$^{2}$Department of Physics, Hubei University, Wuhan 430062, China\\
$^{3}$Department of Physics, The University of Texas at Austin, Austin,
Texas 78712, USA}

\begin{abstract}
The solutions for the helical edge states for an effective continuum model
for the quantum spin Hall effect in HgTe/CdTe quantum wells are presented.
For a sample of a large size, the solution gives the linear dispersion for
the edge states. However, in a finite strip geometry, the edge states at two
sides will couple with each other, which leads to a finite energy gap in the
spectra. The gap decays in an exponential law of the width of sample. The
magnetic field dependence of the edge states illustrates the difference of
the edge states from those of a conventional quantum Hall strip of
two-dimensional electron gas.
\end{abstract}

\maketitle

Recent discovery of quantum spin Hall (QSH) effect brings the Hall family a
new member, an insulator with topological properties of electron bands
distinct from the conventional ones \cite{Day08}. Kane and Mele \cite%
{Kane05PRL} proposed that newly synthesized graphene may exhibit a QSH
effect. This novel state consists of the helical edge states whose
dispersions locate inside the bulk insulating gap. They indicated that the $%
Z_{2}$ topological number distinguishes between the QSH and usual insulating
phases. Benevig \textit{et al} \cite{Bernevig06SCI} realized that the
HgTe/CdTe quantum well is a potential candidate for observing this effect
based on the fact that the electron bands changes from the normal to an
"inverted" type at a critical thickness of the quantum well, which was
confirmed experimentally within one year.\cite{Konig07SCI} Several other
candidates were also proposed, such as GaAs with shear strain \cite%
{Bernevig06PRL} and a multilayer Bi thin film\cite{Murakami06PRL}. Up to
now, considerable efforts have been done to explore the properties of the
QSH effect and topological insulators \cite{Others}. One of the striking
properties is the crossing linear dispersions of the two edge states in
which the electric currents with different spins flow in opposite
directions. Very recently, it was reported that the massive Dirac particles
exist in the bulk of Bi$_{0.9}$Sb$_{0.1}$, which is a hall mark of higher
dimensional quantum spin Hall insulator.\cite{Hsieh-08nature}

The theory of QSH effect in HgTe/CdTe was based on an effective 4-band model
of the HgTe/CdTe quantum well that participates in the inversion crossing of
electron and hole bands derived by Benervig, Hughes and Zhang \cite%
{Bernevig06SCI}. Until present almost all works are based on numerical
solutions of the model in a tight-binding method \cite%
{Kane05PRL,Qi06PRB,Konig08XXX}. It was argued that the QSH insulator state
has gapless and linear dispersions locating inside the bulk insulating gap.
In the present Letter, we present the solution for the edge states of the
4-band model. In a finite strip geometry, linear dispersions for four edge
states are reproduced for a sufficient large sample comparing with the space
distribution scale of the edge states. However, when the size of sample is
comparable with the size scale of the edge state, the edge states near the
crossing point will couple with each other and open a finite energy gap. The
gap decays in an exponential law with the size of the sample. This
demonstrates that the edge states of QSH effect are quite different from the
edge states of quantum Hall (QH) effect.

Here we start from the effective 4-band model for HgTe/CdTe quantum wells 
\cite{Bernevig06SCI} 
\begin{equation}
\mathcal{H}\left( k_{x},k_{y}\right) =\left( 
\begin{array}{cc}
H\left( k\right) & 0 \\ 
0 & H^{\ast }\left( -k\right)%
\end{array}%
\right) ,  \label{H}
\end{equation}%
where $H\left( k\right) =\epsilon _{k}\mathbf{I}_{2}+d^{a}\left( k\right)
\sigma ^{a}$, with $\mathbf{I}_{2}$ being a $2\times 2$ unit matrix, $\sigma
_{a}$ the Pauli matrices, $\epsilon _{k}=C-D\left(
k_{x}^{2}+k_{y}^{2}\right) $, $d^{1}=Ak_{x}$, $d^{2}=Ak_{y}$, and $d^{3}=%
\mathcal{M}\left( k\right) =M-B\left( k_{x}^{2}+k_{y}^{2}\right) $. $A$, $B$%
, $C$, $D$, and $M$ are material specific parameters that are functions of
the thickness of the quantum well. The model is derived from the Kane model
near the $\Gamma $ point in a heterojunction of HgTe/CdTe, and its validity
is limited for small values of $k_{x,y}$. The most striking property of this
system is that the mass or gap parameter $M$ changes sign when the thickness 
$d$ of the quantum well is varied through a critical thickness $d_{c}$ ($%
=6.3 $ nm) associating with the transition of electronic band structure from
a normal to an "inverted" type \cite{Jeschke00}. We will now solve this
continuum model (\ref{H}) in a finite strip geometry of the width $L$ with
the periodic boundary condition in the $x$ direction and an open boundary
condition in the $y$ direction. In this case, $k_{x}$ is a good quantum
number but $k_{y}$ is replaced by using the Peierls substitution $%
k_{y}=-i\partial /\partial y=-i\partial _{y}$. The Hamiltonian (\ref{H}) is
block-diagonal, and the upper $\hat{H}_{\uparrow }$ ($=H\left( k\right) $)
and lower $\hat{H}_{\downarrow }$ ($=H^{\ast }\left( -k\right) $) blocks
describe the states of spin-up and spin-down, respectively. The eigenvalue
problem of the upper and lower blocks can be solved separately, i.e., $\hat{H%
}_{\uparrow }\Psi _{\uparrow }=E\Psi _{\uparrow }$ and $\hat{H}_{\downarrow
}\Psi _{\downarrow }=E\Psi _{\downarrow }$ with $\Psi _{\uparrow }\left(
k_{x},y\right) =e^{ik_{x}x}\left( \psi _{1}\left( k_{x},y\right) ,\psi
_{2}\left( k_{x},y\right) \right) ^{T}$ and $\Psi _{\downarrow }\left(
k_{x},y\right) =e^{ik_{x}x}\left( \psi _{3}\left( k_{x},y\right) ,\psi
_{4}\left( k_{x},y\right) \right) ^{T}$ where $T$ means "transpose". Because
the lower block of the Hamiltonian is the time reversal of the upper block
of the Hamiltonian, the solution $\Psi _{\downarrow }\left( k_{x},y\right)
=\Theta \Psi _{\uparrow }\left( k_{x},y\right) $, where $\Theta =-i\sigma
_{y}K$ is a "time-reversal" operator and $K$ stands for complex conjugation.
Thus we can only focus on the solution for the upper block of this
Hamiltonian.

The set of the eigenvalue equations for the upper block is expressed as 
\begin{eqnarray}
\left[ M-B_{+}\left( k_{x}^{2}-\partial _{y}^{2}\right) \right] \psi
_{1}+A\left( k_{x}-\partial _{y}\right) \psi _{2} &=&E\psi _{1},  \label{HK1}
\\
A\left( k_{x}+\partial _{y}\right) \psi _{1}-\left[ M-B_{-}\left(
k_{x}^{2}-\partial _{y}^{2}\right) \right] \psi _{2} &=&E\psi _{2},
\label{HK2}
\end{eqnarray}%
with $B_{\pm }=B\pm D$. From eqs. (\ref{HK1}) and (\ref{HK2}), using the
trial function $\psi _{1,2}=e^{\lambda y}$, the characteristic equation
gives four roots $\pm \lambda _{1}$ and $\pm \lambda _{2}$, 
\begin{equation}
\lambda _{1,2}^{2}=k_{x}^{2}+F\pm \sqrt{F^{2}-(M^{2}-E^{2})/B_{+}B_{-}}.
\label{LK}
\end{equation}%
where $F=\frac{A^{2}-2\left( MB+ED\right) }{2B_{+}B_{-}}.$ With the boundary
conditions of $\Psi _{\uparrow }\left( k_{x},y=\pm L/2\right) =0$, we have
an analytical expression for the wave function $\Psi _{\uparrow }$ 
\begin{eqnarray}
\psi _{1} &=&\tilde{c}_{+}f_{+}\left( k_{x},y\right) +\tilde{c}%
_{-}f_{-}\left( k_{x},y\right) ,  \label{W1} \\
\psi _{2} &=&\tilde{d}_{+}f_{+}\left( k_{x},y\right) +\tilde{d}%
_{-}f_{-}\left( k_{x},y\right) ,  \label{W2}
\end{eqnarray}%
with 
\begin{eqnarray}
f_{+}\left( k_{x},y\right)  &=&\frac{\cosh \left( \lambda _{1}y\right) }{%
\cosh \left( \lambda _{1}L/2\right) }-\frac{\cosh \left( \lambda
_{2}y\right) }{\cosh \left( \lambda _{2}L/2\right) }, \\
f_{-}\left( k_{x},y\right)  &=&\frac{\sinh \left( \lambda _{1}y\right) }{%
\sinh \left( \lambda _{1}L/2\right) }-\frac{\sinh \left( \lambda
_{2}y\right) }{\sinh \left( \lambda _{2}L/2\right) }.
\end{eqnarray}%
The non-trivial solution for the coefficients $\tilde{c}_{\pm }$ and $\tilde{%
d}_{\pm }$ in the wave functions leads to a secular equation 
\begin{equation}
\frac{\tanh \frac{\lambda _{1}L}{2}}{\tanh \frac{\lambda _{2}L}{2}}+\frac{%
\tanh \frac{\lambda _{2}L}{2}}{\tanh \frac{\lambda _{1}L}{2}}=\frac{\alpha
^{2}\lambda _{2}^{2}+\beta ^{2}\lambda _{1}^{2}-k_{x}^{2}\left( \alpha
-\beta \right) ^{2}}{\alpha \beta \lambda _{1}\lambda _{2}},  \label{EK}
\end{equation}%
where $\alpha (E)=E-M+B_{+}k_{x}^{2}-B_{+}\lambda _{1}^{2}$, and $\beta
(E)=E-M+B_{+}k_{x}^{2}-B_{+}\lambda _{2}^{2}$. Eqs. (\ref{LK}) and (\ref{EK}%
) determine the energy spectra of the upper block $\hat{H}_{\uparrow }$.

Let us first consider the general properties of the solution for $\lambda
_{1,2}$. In a large $L$ limit, a purely imaginary $\lambda =ik_{y}$ is
always the solutions of the equation, and gives two branches of spectra, $%
E_{\pm }=\epsilon _{k}\pm \sqrt{\left( M-Bk^{2}\right) ^{2}+A^{2}k^{2}}$.
These are the bulk spectra and revised by the $A$-term slightly, and
corresponding solutions span in the whole space. Except for these imaginary
solutions, there also exist real solutions when 
\begin{equation}
A^{2}/B_{+}B_{-}>\max \{2M/B,4M/B\}.
\end{equation}%
In the large $L$ limit, $\tanh \left( \lambda _{1,2}L/2\right) =1,$ and Eq. (%
\ref{EK}) gives 
\begin{equation}
E_{\pm }=M-B_{+}\lambda _{1}\lambda _{2}\pm B_{+}(\lambda _{1}+\lambda
_{2})k_{x}-B_{+}k_{x}^{2}.
\end{equation}%
As the $\lambda _{1,2}$ approaches to constant near $k_{x}=0$, $E_{\pm
}\left( k_{x}\right) \simeq -MD/B\pm A\sqrt{B_{+}B_{-}/B^{2}}k_{x}$. For
real roots $\lambda _{1,2}$, the function $f_{\pm }(y)$ are distributed
dominantly near the edge ($y=\pm L/2$) in the scale of $\lambda _{1,2}^{-1}$%
. This result is consistent with those by means of the tight binding
approximation \cite{Qi06PRB}.

\begin{figure}[tbp]
\includegraphics[width=7.5cm]{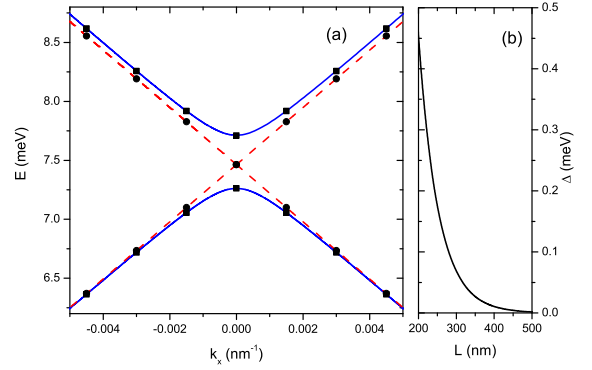}
\caption{{}(Color online) (a) Energy spectra of edge states for $L=200$ nm
(blue solid lines) and $L=1000$ nm (red dashed lines) by solving eqs. (%
\protect\ref{LK}) and (\protect\ref{EK}) for the HgTe/CdTe quantum well
thickness $d=7.0$ nm. The parameters are given in the text. As a comparison,
numerical results of tight binding approximation, which was used in Ref.%
\protect\cite{Qi06PRB} for the same parameters are also plotted as black
squares for $L=200$ nm and black dots for $L=1000$ nm. (b) The width
dependence of the energy gap $\Delta $ for a finite width strip of HgTe/CdTe
quantum well of thickness $d=7.0$ nm. Specifically, $\Delta (L=200$ nm$%
)=0.4509$ meV, $\Delta (500$ nm$)=1.6\times 10^{-3}$ meV, and $\Delta
(L=1000 $ nm$)=1.41\times 10^{-7}$ meV.}
\end{figure}

In the following we shall concentrate on the real solution of $\lambda $
with a finite $L$, i.e. the solutions for the edge states. For real $\lambda 
$ and finite $L$, the right hand side of Eq. (\ref{EK}) is always greater
than $2$. If $\lambda _{1,2}L>>1$, it is approximately $2+4e^{-2\lambda
_{2}L}$ (assuming $\lambda _{1}>>\lambda _{2}$). From Eq. (\ref{EK}), it is
found that a finite energy gap $\Delta =E_{+}-E_{-}$ opens at $k_{x}=0$%
\begin{equation}
\Delta \simeq \frac{4\left\vert AB_{+}B_{-}M\right\vert }{\sqrt{B^{3}\left(
A^{2}B-4B_{+}B_{-}M\right) }}e^{-\lambda _{2}L},  \label{gap}
\end{equation}%
which decays in an exponential law of $L$. This is the main consequence in
the present work.

In general cases, we have numerical solution of the equations. As a concrete
example, we adopt the parameters for the inverted HgTe/CdTe quantum well of
thickness $d=7.0$ nm from the reference \cite{Konig08XXX} for all numerical
calculations in the present Letter: $A=364.5$ meV nm, $B=-686$ meV nm$^{2}$, 
$M=-10$ meV, $D=-512$ meV nm$^{2}$. For $L=1000$ nm, one has $\lambda
_{1}=0.7797$ nm$^{-1}$ and $\lambda _{2}=0.0187$ nm$^{-1}$ at $k_{x}=0$. The
energy gap is very tiny, $\Delta =E_{+}-E_{-}=1.41\times 10^{-7}$ meV.
However, for $L=200$ nm at $k_{x}=0$, the gap $\Delta =0.4509$ meV, which
becomes large enough to be measurable in experiments. We plot the energy
spectra for the edge states of several sizes in Fig. 1(a). However, for a
narrow width $L$ (e.g., $L=200$ nm), there is the parabolic-like spectrum
near $k_{x}=0$. The size-dependence of the energy gap is plotted in Fig.
1(b).

The corresponding wave functions of eigenvalues $E_{\pm }\left( k_{x}\right) 
$ yields 
\begin{eqnarray}
\Psi _{\uparrow +} &=&\tilde{c}_{+}e^{ik_{x}x}\left( f_{+}+\gamma
_{k_{x}}^{+}f_{-},\eta _{1}^{+}f_{-}+\gamma _{k_{x}}^{+}\eta
_{2}^{+}f_{+}\right) ^{T},  \label{F1} \\
\Psi _{\uparrow -} &=&\tilde{c}_{-}e^{ik_{x}x}\left( f_{-}+\gamma
_{k_{x}}^{-}f_{+},\eta _{2}^{-}f_{+}+\gamma _{k_{x}}^{-}\eta
_{1}^{-}f_{-}\right) ^{T},  \label{F2}
\end{eqnarray}%
where%
\begin{eqnarray*}
\eta _{1}^{\pm } &=&\left. \frac{B_{+}\left( \lambda _{1}^{2}-\lambda
_{2}^{2}\right) /A}{\lambda _{1}\coth \frac{\lambda _{1}L}{2}-\lambda
_{2}\coth \frac{\lambda _{2}L}{2}}\right\vert _{E=E_{\pm }}, \\
\eta _{2}^{\pm } &=&\left. \frac{B_{+}\left( \lambda _{1}^{2}-\lambda
_{2}^{2}\right) /A}{\lambda _{1}\tanh \frac{\lambda _{1}L}{2}-\lambda
_{2}\tanh \frac{\lambda _{2}L}{2}}\right\vert _{E=E_{\pm }}, \\
\gamma _{k_{x}}^{+} &=&\left. \frac{B_{+}\left( \lambda _{1}^{2}-\lambda
_{2}^{2}\right) k_{x}\eta _{1}/\eta _{2}}{\beta \lambda _{1}\tanh \frac{%
\lambda _{1}L}{2}-\alpha \lambda _{2}\tanh \frac{\lambda _{2}L}{2}}%
\right\vert _{E=E_{+}}, \\
\gamma _{k_{x}}^{-} &=&\left. \frac{B_{+}\left( \lambda _{1}^{2}-\lambda
_{2}^{2}\right) k_{x}\eta _{2}/\eta _{1}}{\beta \lambda _{1}\coth \frac{%
\lambda _{1}L}{2}-\alpha \lambda _{2}\coth \frac{\lambda _{2}L}{2}}%
\right\vert _{E=E_{-}}.
\end{eqnarray*}%
and $\tilde{c}_{\pm }$ are normalization constants. The solutions can be
simplified in the limit of large $L$ as $\eta _{1}^{\pm }=\eta _{2}^{\pm
}=\eta =B_{+}(\lambda _{1}+\lambda _{2})/A$ and $\gamma _{k_{x}}^{+}=-\gamma
_{k_{x}}^{-}=-$sgn$(k_{x})$. The other two solutions for the lower block can
be produced by means of the time reversal operation, $\Psi _{\downarrow \pm
}=\Theta \Psi _{\uparrow \pm }$, and the spectra are degenerate with those
of the upper block.

\begin{figure}[tbp]
\includegraphics[width=8cm]{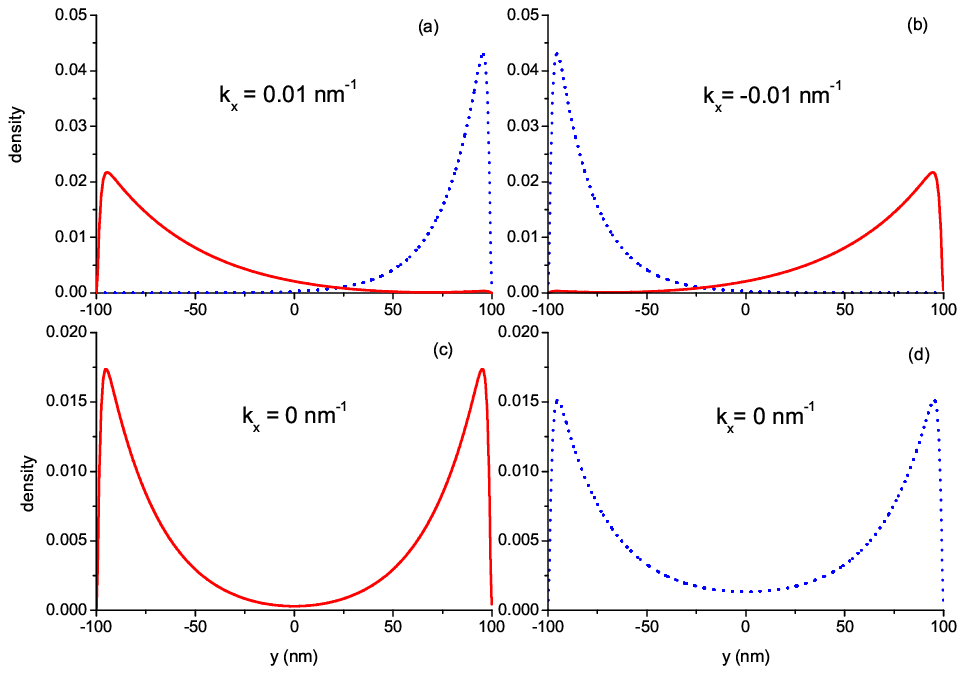}
\caption{{}(Color online) The density distribution of the two edge states $%
\Psi _{\uparrow \pm }\left( k_{x},y\right) $ for $L=200$ nm. (a) The red
solid line corresponds to $\left\vert \Psi _{\uparrow +}\left(
k_{x},y\right) \right\vert ^{2}$ and the blue dotted line to $\left\vert
\Psi _{\uparrow +}\left( -k_{x},y\right) \right\vert ^{2}$ at $k_{x}=0.01$ nm%
$^{-1}$; (b) The red solid line corresponds to $\left\vert \Psi _{\uparrow
-}\left( k_{x},y\right) \right\vert ^{2}$ and the blue dotted line to $%
\left\vert \Psi _{\uparrow -}\left( -k_{x},y\right) \right\vert ^{2}$ at $%
k_{x}=-0.01$ nm$^{-1}$; (c) and (d) for $k_{x}=0$ nm$^{-1}$.}
\label{wave}
\end{figure}

According to the present analytic solutions of wave functions, the density
distribution of the functions are mainly determined by the two length
scales, $\lambda _{1,2}^{-1}$. In the example in Fig. 1, we notice that $%
\lambda _{2}^{-1}>>\lambda _{1}^{-1}$. For a large size of the sample, the
density of the wave function increases in an exponential law in the scale of 
$\lambda _{2}^{-1}$ and then decays exponentially in a scale of $\lambda
_{1}^{-1}$ near the boundaries, which is consistent with the work by K\"{o}%
nig \textit{et al}.\cite{Konig08XXX} The wave function almost vanishes far
away from the boundaries if the width of the sample is much larger than $%
\lambda _{2}^{-1}$. As an example, the density distributions of $\Psi
_{\uparrow \pm }\left( k_{x},y\right) $ for $L=200$ nm are plotted for
demonstration in Fig. 2 where $\lambda _{2}^{-1}=55.9$ nm and $51.8$ nm at $%
k_{x}=0$. The states of $\Psi _{\uparrow +}\left( k_{x},y\right) $ and $\Psi
_{\uparrow -}\left( -k_{x},y\right) $ ($k_{x}>0$) have the same spin ($%
\propto \left( 1,-\eta \right) ^{T}$ in the large $L$ limit) and the
positive velocity, $v_{x}$ ($=+A\sqrt{B_{+}B_{-}/B^{2}}$) $>0$ when $k_{x}$
is far away from $k_{x}=0$ and the density distribution is located at one
side while the states of $\Psi _{\uparrow +}\left( -k_{x},y\right) $ and $%
\Psi _{\uparrow -}\left( +k_{x},y\right) $ ($k_{x}>0$) have another spin ($%
\propto \left( 1,\eta \right) ^{T}$ in the large $L$ limit) and a negative
velocity, $-v_{x}<0$, and are distributed on the other side. From the
solution we found that $\Psi _{\uparrow \pm }\left( k_{x},y\right) $ and $%
\Psi _{\uparrow \pm }\left( -k_{x},y\right) $ couple near $k_{x}=0$ due to
the finite size effect. Consequently, the states with different spins mix
together and the densities of the wave functions $\Psi _{\uparrow \pm
}\left( k_{x}=0,y\right) $ are symmetrically distributed at the two sides.
This fact is consistent with the opening of an energy gap in the spectra at $%
k_{x}=0$. 
\begin{figure}[h]
\includegraphics[width=7.0cm]{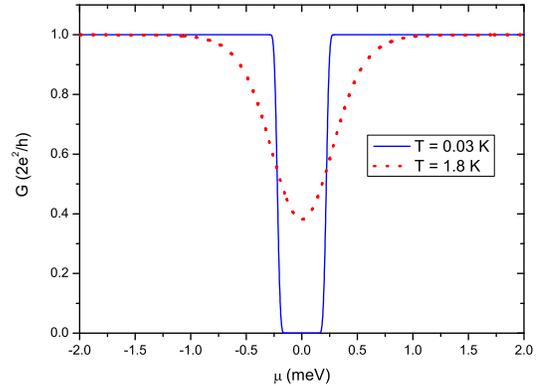}
\caption{{}(Color online) The variations of conductance $G$ via the chemical
potential $\protect\mu $ inside the bulk insulting gap for a HgTe/CdTe
quantum well of $L=200$ nm at temperatures $T=30$ mK and $1.8$ K,
respectively. Note that the energy zero point is shifted to the center of
the gap $\Delta $ ($=0.4509$ meV).}
\label{G}
\end{figure}

The charge conductance of a QSH phase in a strip was predicted theoretically
to be $2e^{2}/h$ due to the presence of two spin-resolved conducting
channels at the edges of the strip, which was observed experimentally in two
samples with sizes of $\left( 1.0\times 1.0\right) $ $\mu $m$^{2}$ and $%
\left( 1.0\times 0.5\right) $ $\mu $m$^{2}$, respectively.\cite{Konig07SCI}
The finite size effect will modify the conductance of the QSH phase.
Following the Landauer-B\"{u}ttiker formula,\cite{DattaBook} the charge
conductance has the form, 
\begin{equation*}
G\left( \Delta \right) =\frac{2e^{2}}{h}\left[ \frac{1}{e^{\left( \frac{%
\Delta }{2}-\mu \right) /k_{B}T}+1}-\frac{1}{e^{\left( -\frac{\Delta }{2}%
-\mu \right) /k_{B}T}+1}+1\right] .
\end{equation*}%
$G\left( \Delta \right) \rightarrow 2e^{2}/h$ at low temperatures only when
the chemical potential locates out of the gap $\Delta (L)$. Below the
temperature of $k_{B}T^{\ast }=\Delta $, a dip will be obviously exhibited.
We plot the temperature dependence of the conductance in Fig. 3. In the
experiment by K\"{o}nig \textit{et al}\cite{Konig07SCI}, the smallest sample
has $L=500$ nm, and the measurement was performed at $30$ mK. The calculated 
$\Delta =1.6\times 10^{-3}$ meV ($19$ mK), which is already comparable with
the experiment temperature. For a smaller example, the energy gap of $L=200$
nm is about $\Delta =0.4509$ meV, and the corresponding temperature is
enhanced to $T^{\ast }=5.22$ K. Experimentally, the data for two smaller
samples are close to the value of $2e^{2}/h$ while the data for the wider
samples obviously deviate from the value. From the present exact solution,
it is believed that the value will be also modified at lower temperatures
for even smaller samples. By the way, it is also worth noting that the gap
is also highly sensitive to the thickness of quantum well, as all parameters
in Eq. (\ref{H}) are functions of the thickness.

\begin{figure}[tbph]
\includegraphics[width=6.5cm]{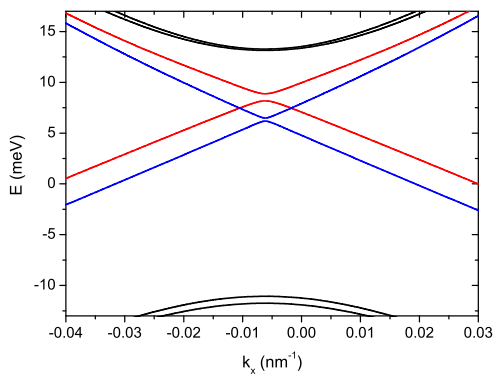}
\caption{{}(Color online) The energy dispersion of the edge states in a weak
field of $B=0.04$T. The blue lines are for the upper block and red lines are
for the lower block. The back lines is for the bulk spectra. The width of
sample is $L=200$nm.}
\end{figure}

Finally, our solution shows that the QSH edge states is quite different from
the edge states of a conventional QH strip. For a QH strip with
translational symmetry along the strip, the states are classified with the
momentum $k_{x}$. The edge states at the two sides have different $k_{x}$.%
\cite{MacDonald} So they do not mix together even when the two states
overlap in the space if there is no other scatterers or interactions in
between two edges. In the case of a QSH strip, the edge states at the two
sides have the same $k_{x}$. Near the anti-crossing points, the two states
has the nearly equal energy and momentum. So they can couple together to
generate an energy gap when their energies becomes closer and the wave
functions have overlaps in a finite space. The magentic field dependence of
the QSH edge states also reflects this peculiar property. Consider the
sample is subjected to a weak perpendicular magnetic field $\mathbf{B}_{z}$.
Using the Peierls substitution, $k_{x}\rightarrow k_{x}-eA_{x}/\hbar $ in
Eq.(1) by taking the gauge, $A_{x}=-B_{z}y$ (for $\left\vert y\right\vert
<L/2$) and $A_{y}=0$ in order to keep $k_{x}$ a good quantum number. As the
wave functions of the edge states decay exponentially, the expectation value
of $A_{x}$ in the two edge states is proportional to $L$ approximately for $%
L>>\lambda _{1,2}^{-1}$. The energy will be shifted by $\Delta E=+g\mu
_{B}B_{z}$ ($g\approx m_{e}v_{x}\left[ L-\lambda _{1}^{-1}-\lambda
_{2}^{-2}-2(\lambda _{1}+\lambda _{2})^{-1}\right] /\hbar $) near $k_{x}=0$.
Thus the energy spectra of the two edge states of the upper block will shift
downward or upward $E(k_{x})\approx \pm v_{x}\hbar k_{x}+g\mu _{B}B_{z}$ for
a large $L$. By increasing the magnetic field $B_{z}$ the anti-crossing
point of energy spectra is eventually moved out of the bulk insulating gap,
and the spectra will not crossing in momentum between the gap. However,
another two branches of the spectra of the lower block will move in an
opposite direction, and the two sets of the spectra may cross in momentum
inside the insulating gap. A more detailed calculation is to use the exact
solutions of Eqs. (13) and (14) at $B_{z}=0$ as a basis to truncate the
model of Eq.(1) in a $B_{z}$ field to an effective one. Numerical results
are plotted in Fig.4. The energy shift is very sensitive to the magnetic
field and the width of the sample. As for the conductance, the value of $%
2e^{2}/h$ will recover near the crossing points. Thus the magnetoresistance
is very sensitive to a tiny field, which might have potential application
for a sensitive detection.

SQS thanks S. C. Zhang for helpful discussions. This work was supported by
the Research Grant Council of Hong Kong under Grant No.: HKU 7042/06P, and
the CRCG of the University of Hong Kong.


\begin{thebibliography}{99}
\bibitem{Day08} C. Day, Physics Today \textbf{61}, 19 (2008).

\bibitem{Kane05PRL} C. L. Kane and E. J. Mele, Phys. Rev. Lett. \textbf{95},
146802 (2005); \textit{ibid}. \textbf{95}, 226801 (2005).

\bibitem{Bernevig06SCI} B. A. Bernevig \textit{et al.}, Science \textbf{314}%
, 1757 (2006).

\bibitem{Konig07SCI} M. K\"{o}nig \textit{et al.}, Science \textbf{318}, 766
(2007).

\bibitem{Bernevig06PRL} B. A. Bernevig and S. C. Zhang, Phys. Rev. Lett. 
\textbf{96}, 106802 (2006).

\bibitem{Murakami06PRL} S. Murakami, Phys. Rev. Lett. \textbf{97}, 236805
(2006);

\bibitem{Others} D. N. Sheng \textit{et al.}, Phys. Rev. Lett. \textbf{97},
036808 (2006); C. Wu, B. A. Bernevig, and S. C. Zhang, \textit{ibid.} 
\textbf{96}, 106401 (2006); M. Onoda \textit{et al.}, \textit{ibid.} \textbf{%
98}, 076802 (2007); C. Xu and J. E. Moore, Phys. Rev. B \textbf{73}, 045322
(2006); L. Fu and C. L. Kane, \textit{ibid.} \textbf{74}, 195312 (2006); T.
Fukui and Y. Hatsugai, \textit{ibid.} \textbf{75}, 121403 (2007); H. Obuse 
\textit{et al.}, \textit{ibid.} \textbf{76}, 075301 (2007); S. Murakami, S.
Iso, Y. Avishai, M. Onoda, and N. Nagaosa, \textit{ibid.} \textbf{76},
205304 (2007); X. Dai \textit{et al.}, \textit{ibid.} \textbf{77}, 125319
(2008); Z. H. Qiao \textit{et al.}, arXiv: 0711.1005.

\bibitem{Hsieh-08nature} D. Hsieh \textit{et al.}, Nature 452, 970 (2008).

\bibitem{Qi06PRB} X. L. Qi \textit{et al.}, Phys. Rev. B \textbf{74}, 085308
(2006).

\bibitem{Konig08XXX} M. K\"{o}nig \textit{et al.}, J. Phys. Soc. Jpn 77,
031007 (2008).

\bibitem{Jeschke00} A. Pfeuffer-Jeschke, thesis, University of W\"{u}rzburg
(2000).

\bibitem{DattaBook} S. Datta, \textit{Electronic transport in mesoscopic
systems}, (Cambridge University Press, Cambridge, 1995).

\bibitem{MacDonald} B. I. Halperin, Phys. Rev. B 25, 2185 (1982); A. H.
MacDonald and P. Steda, Phys. Rev. B 29, 1616 (1984).
\end{thebibliography}
\end{document}